\documentclass[journal]{vgtc}                

\ifpdf
  \pdfoutput=1\relax                   
  \usepackage{graphicx}                
  \DeclareGraphicsExtensions{.pdf,.png,.jpg,.jpeg} 
\else
  \usepackage{graphicx}                
  \DeclareGraphicsExtensions{.eps}     
\fi%

\graphicspath{{figures/}{pictures/}{images/}{./}} 

\usepackage{microtype}                 
\PassOptionsToPackage{warn}{textcomp}  
\usepackage{textcomp}                  
\usepackage{mathptmx}                  
\usepackage{times}                     
\usepackage{cite}                      
\usepackage{tabu}                      
\usepackage{booktabs}                  

\usepackage{amsmath}
\usepackage{amssymb}
\mathchardef\mhyphen="2D 

\usepackage{mathtools}
\usepackage{listings}
\usepackage{xspace}

\usepackage{multirow}
\usepackage{textcomp}

\usepackage{enumitem}

\usepackage{hyperref}

\newcommand{\etc}{{\em etc}\xspace}
\newcommand{\code}[1]{\texttt{#1}}
\newcommand{\eg}{{\em e.g.,}\xspace}

\newcommand{\q}[1]{\emph{``#1''}}

\newcommand{\codeblockOneOrange}{\raisebox{-.2\baselineskip}{\includegraphics[height=0.9\baselineskip, keepaspectratio]{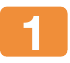}}}
\newcommand{\codeblockTwoOrange}{\raisebox{-.2\baselineskip}{\includegraphics[height=0.9\baselineskip, keepaspectratio]{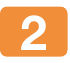}}}

\newcommand{\codeblockOneDotOneBlue}{\raisebox{-.2\baselineskip}{\includegraphics[height=0.9\baselineskip, keepaspectratio]{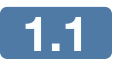}}}

\newcommand{\codeblockTwoDotOneBlue}{\raisebox{-.2\baselineskip}{\includegraphics[height=0.9\baselineskip, keepaspectratio]{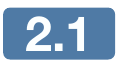}}}

\newcommand{\codeblockOne}{\raisebox{-.2\baselineskip}{\includegraphics[height=0.9\baselineskip, keepaspectratio]{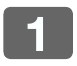}}}
\newcommand{\codeblockOneDotOne}{\raisebox{-.2\baselineskip}{\includegraphics[height=0.9\baselineskip, keepaspectratio]{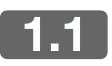}}}
\newcommand{\codeblockOneDotTwo}{\raisebox{-.2\baselineskip}{\includegraphics[height=0.9\baselineskip, keepaspectratio]{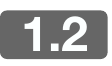}}}
\newcommand{\codeblockOneDotThree}{\raisebox{-.2\baselineskip}{\includegraphics[height=0.9\baselineskip, keepaspectratio]{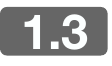}}}
\newcommand{\codeblockTwo}{\raisebox{-.2\baselineskip}{\includegraphics[height=0.9\baselineskip, keepaspectratio]{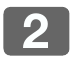}}}

\newcommand{\codeblockThree}{\raisebox{-.2\baselineskip}{\includegraphics[height=0.9\baselineskip, keepaspectratio]{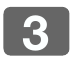}}}

\newcommand{\codeblockMs}{\raisebox{-.2\baselineskip}{\includegraphics[height=0.9\baselineskip, keepaspectratio]{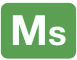}}}
\newcommand{\codeblockMd}{\raisebox{-.2\baselineskip}{\includegraphics[height=0.9\baselineskip, keepaspectratio]{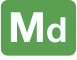}}}
\newcommand{\codeblockAs}{\raisebox{-.2\baselineskip}{\includegraphics[height=0.9\baselineskip, keepaspectratio]{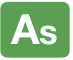}}}

\newcommand{\designA}{\raisebox{-.2\baselineskip}{\includegraphics[height=0.9\baselineskip, keepaspectratio]{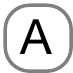}}}
\newcommand{\designB}{\raisebox{-.2\baselineskip}{\includegraphics[height=0.9\baselineskip, keepaspectratio]{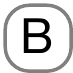}}}

\onlineid{0}

\vgtccategory{Research}
\vgtcpapertype{system}

\title{Gemini: A Grammar and Recommender System for Animated Transitions in Statistical Graphics}

\author{Younghoon Kim and Jeffrey Heer}
\authorfooter{
\item
 Younghoon Kim and Jeffrey Heer are with the University of Washington. E-mails: yhkim01, jheer@uw.edu
}

\shortauthortitle{Kim & Heer: Gemini: A Grammar and Recommender System for Animated Transitions in Statistical Graphics}

 \abstract{Animated transitions help viewers follow changes between related visualizations. Specifying effective animations demands significant effort: authors must select the elements and properties to animate, provide transition parameters, and coordinate the timing of stages. To facilitate this process, we present Gemini, a declarative grammar and recommendation system for animated transitions between single-view statistical graphics. Gemini specifications define transition ``steps'' in terms of high-level visual components (marks, axes, legends) and composition rules to synchronize and concatenate steps. With this grammar, Gemini can recommend animation designs to augment and accelerate designers’ work. Gemini enumerates staged animation designs for given start and end states, and ranks those designs using a cost function informed by prior perceptual studies. To evaluate Gemini, we conduct both a formative study on Mechanical Turk to assess and tune our ranking function, and a summative study in which 8 experienced visualization developers implement animations in D3 that we then compare to Gemini’s suggestions. We find that most designs (9/11) are exactly replicable in Gemini, with many (8/11) achievable via edits to suggestions, and that Gemini suggestions avoid multiple participant errors.}


\CCScatlist{ 
 \CCScat{K.6.1}{Management of Computing and Information Systems}%
{Project and People Management}{Life Cycle};
 \CCScat{K.7.m}{The Computing Profession}{Miscellaneous}{Ethics}
}

\teaser{
  \centering
  \includegraphics[width=\linewidth]{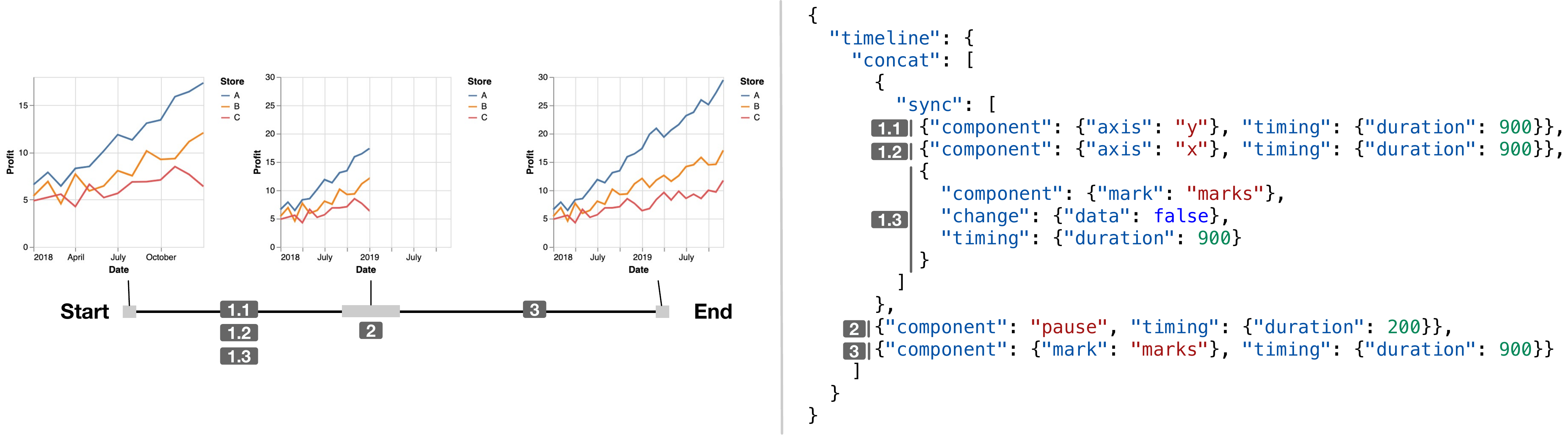}
  \vspace{-20pt}
  \caption{Left: An animated transition zooms a line chart to a larger time window. The timeline indicates the animation sequence; the enlarged gray interval indicates a pause. Right: The Gemini specification for the transition, which changes the scales of the axes and lines (\protect\codeblockOne{}), pauses (\protect\codeblockTwo{}), and then extends the lines (\protect\codeblockThree). The numbered items represent \emph{steps}, the basic units of the Gemini grammar.}
    \label{fig:eg1_line}
}

\vgtcinsertpkg

\begin{document}


\firstsection{Introduction}

\maketitle

When exploring data or communicating results, people often transition between related statistical graphics. To facilitate understanding of what has changed across a transition, visualization researchers have developed and studied animation techniques. Prior studies have examined the effectiveness of animation for conveying transitions~\cite{cone_tree, anim_mental_map, anim_decision_making, hops, hops_trend, face2face} and proposed guidelines and strategies for animation design, including the use of techniques such as staging and staggering~\cite{anim_transition, anim_trend_vis, tversky, agg_anim, staggering}.

Despite this guidance, creating effective animations remains challenging, as current tools either cannot express more nuanced designs~\cite{dataclip, gganimate} or do so only with significant effort. A designer may need to select elements and properties to animate, specify transition parameters, and coordinate the relative timing of separate stages. 
Using D3~\cite{d3}, for example, the implementation of animated transitions often requires manual orchestration of animation stages using a transition abstraction that intertwines visual encoding and animation specifications, impeding rapid design exploration and reuse.

To reduce this hurdle we contribute Gemini, a declarative grammar and recommender system for animated transitions between two single-view statistical graphics. In Gemini, animated transitions are formally represented by \textit{transition steps} in terms of high-level visual components (marks, axes, legends) and \textit{composition rules} synchronizing and concatenating steps into staged animations (\autoref{fig:eg1_line}). This formal representation allows software to reason systematically about animated transitions. By taking advantage of this formalism, Gemini's recommender system produces candidate animated transitions between given start and end visualization states expressed in the Vega visualization grammar~\cite{vega}. The recommendations can facilitate the design process by serving as starting points so that users need not manually create animations from scratch. 



We begin by articulating our design goals for the Gemini grammar. We target a balance between expressiveness and ease-of-use by reviewing existing animation tools and alternate approaches. We also observe how people describe animated transitions in a preliminary study to gauge a proper level of abstraction. We then introduce the primitives of Gemini along with our backing design rationale.

We go on to describe the workflow of the Gemini recommender system: change detection, enumeration, and ranking. We introduce a heuristic cost function that ranks enumerated candidate designs based on their complexity. To assess and refine the cost function, we conduct a user study and tune the cost function parameters to user preferences by promoting single-stage designs and demoting multi-stage ones.

We verify the utility of Gemini by replicating animated transitions created by designers using D3. We observe that among 11 user-crafted animated transitions, 9 can be expressed exactly in the Gemini grammar, 5 can be replicated by changing only timing parameters of a top-3 Gemini suggestion, and a total of 8 can be achieved by editing a top suggestion.
We also find that 7/11 designs exhibit mistakes that all Gemini-produced designs avoid. These findings show Gemini's potential to suggest useful starting points for user-desired animations. We conclude by discussing ways to improve Gemini's user interface, achieve more nuanced suggestions, enhance expressiveness, and overcome implementation challenges in our current proof-of-concept system.

\section{Related Work}

Animated transitions are used to convey state changes and engage viewers. We focus on animated transitions between statistical graphics, with the goal of accurately conveying changes, directing attention, and helping viewers stay oriented.

\subsection{Animated Transitions}

Animation is a common method for conveying changes between visualization states. Prior research has found that animation can help viewers build mental maps of spatial information~\cite{anim_mental_map}, make decisions~\cite{anim_decision_making}, and remain oriented across transitions~\cite{cone_tree, polyarchy}. More recently, researchers found that animations can outperform small multiple encodings in a comparison task~\cite{face2face} and outperform static aggregate uncertainty visualizations in value judgment and trend inference tasks~\cite{hops, hops_trend}.

However, the effectiveness of animation has also been the subject of skepticism. Tversky et al.~\cite{tversky} scrutinized studies showing advantages for animation over static transitions and found that the animation conditions conveyed more information than the static conditions. They proposed two high-level principles, \emph{congruence} and \emph{apprehension}, for animation design. Robertson et al.~\cite{anim_trend_vis} found that animation was less efficient than static small multiple charts for time-series data analysis, but was preferred by users in a presentation context. In addition, Hullman et al.~\cite{chart_junk} identified conditions where animation may be preferable to static representations.
Heer \& Robertson~\cite{anim_transition} contribute strategies for achieving Tversky et al.'s principles in the context of statistical graphics. They compared the performance of animated and static transitions for object tracking and value estimation tasks, finding significant advantages for animated transitions.


Researchers have also categorized different types of transitions between statistical graphics. Heer \& Robertson~\cite{anim_transition} propose seven categories, such as \emph{view transformation}, \emph{substrate transformation}, \etc{}. Kim et al.~\cite{graphscape} identify atomic editing operations that can be combined to define transitions. We use these prior taxonomic treatments to guide the design of Gemini and establish a necessary expressive gamut.

Specific animation techniques have also been investigated. In prior studies, participants often preferred staged animations (which break animations into separate parts) over direct interpolation, although not always with significant task performance benefits~\cite{anim_transition, agg_anim}. Staggering techniques, which individually delay visual elements to decrease occlusion, have also been examined, with no significant impact on object tracking performance~\cite{staggering}. Among temporal distortion (pacing) strategies, slow-in slow-out outperformed others in an object tracking task~\cite{siso}. Gemini supports these techniques so that users can explore a broad spectrum of alternatives.

The perception of data point trajectories might also be improved using bundling~\cite{trajectory} and vector fields~\cite{vector_field}. However, Gemini does not currently support trajectory-related strategies, as they require low-level specifications (spatial interpolation functions of mark elements) that are not applicable to transitions other than those between scatter plots.

\subsection{Event Structure In Perceptual Psychology}

Zacks \& Tversky review how people conceptualize events in perception~\cite{event-structure}. They maintain that events are structured in two hierarchical ways: \textit{partonomy} (one event can be a part of another bigger event) and \textit{taxonomy} (one event is an instance of one category). For example, in event partonomies, two event segments of ``changing the color of visual marks'' and ``filtering out the marks'' are part of ``the change of the marks.'' In event taxonomies, the former is a kind of ``encoding change'' and the latter a kind of ``data transformation.'' On top of these hierarchical structures, people show better perception when communicating at \textit{a basic level} of the event segments. To align with this psychological framing, we determined the basic level of events in Gemini (``steps'') by conducting preliminary interviews. In addition, Gemini arranges these events in a hierarchical structure (timeline or block) to support event segmentation and programmatic enumeration. 

\subsection{Declarative Grammars for Visualization}

Declarative grammars for specifying visualizations~\cite{protovis, vega-lite, vega, polaris, ggplot2} provide several benefits by letting users think about ``what'' to visualize rather than ``how'' to implement it. First, users can explore more visual designs if the required implementation effort is reduced. In addition, users can create more robust or scalable designs (e.g., dealing with data size or supporting multiple platforms) because control flow and execution can be synthesized and optimized by a compiler. Most importantly, a declarative grammar can provide a representation for programmatic enumeration and search over a design space~\cite{graphscape, voyager}. These benefits similarly inspire Gemini. While Gemini uses start and end visualizations specified using the Vega grammar, its approach could be readily applied to other tools (\eg{} ggplot2~\cite{ggplot2}) that use visualization primitives based on Wilkinson's \emph{The Grammar of Graphics}~\cite{gog}: abstractions of data, visual marks, encodings, and guide elements.

Every design tool must make trade-offs between \emph{expressiveness} and \emph{ease-of-use}. Focusing on animated transitions, DataClip~\cite{dataclip} uses a typology of animation designs to help non-experts assemble transition ``clips.'' It prioritizes ease-of-use but constrains expressiveness through the use of pre-defined types and parameters. In contrast, D3~\cite{d3} targets maximal expressiveness so that experts can create novel designs. D3 offers considerable control but requires that users master its \emph{transition} API and manage stages. In an intermediate position is gganimate~\cite{gganimate}, which lets users animate ggplot2~\cite{ggplot2} charts by appending animation directives to the original chart specification.
This approach is naturally limited to transitions among states represented by a \emph{single} visualization specification, such as underlying data changes.

Most recently, Ge et al.~introduce Canis~\cite{canis}, a declarative grammar for animating dSVG-formatted charts. Canis uses low-level selections using W3C Selector syntax~\cite{selectorAPI} (similar to D3) and supports non-hierarchical timeline compositions using constraints such as  \emph{``start with previous''} and \emph{``start after previous.''} Critically, Canis does not perform automatic reasoning and recommendation over transition designs.

With Gemini, we target a different balance of expressiveness and ease-of-use, supporting a wider gamut of designs, including multi-stage transitions between diverse start and end states.
We also target a higher-level specification than D3 or Canis, intended to correspond more intuitively with the perceptual structure of the animated transition. By doing so, Gemini helps both computer systems and developers directly refer to and configure perceptually salient units in the specifications. 

\subsection{Automated Visualization Design}

APT~\cite{APT} automates visualization design using importance-ranked input data fields. It enumerates designs by assigning more effective visual attributes to higher priority data fields while ruling out fields that violate expressiveness rules. Other visualization systems employ heuristics~\cite{showme} or hand-tuned score functions~\cite{voyager, voyager2} to recommend visualization designs. In sequencing a set of visualizations, GraphScape~\cite{graphscape} uses a hybrid approach that relies on linear programming to calculate the weights of edit operations from heuristic constraints. More recently, Moritz et al.'s Draco system~\cite{draco} adapts the scoring function based on a given knowledge base; Lin et al. extend this work to Dziban~\cite{dziban}, a system that uses the context of prior visualizations to make suggestions aligned with users' expectations. Gemini provides a base representation enabling the programmatic enumeration and evaluation of animated transitions. We illustrate Gemini’s recommendation feature using a heuristic scoring function and discuss ways to support advanced recommendation tactics suggested by these prior works.

\section{Preliminary Interviews on Animation Design}

To observe how people segment animated transitions and to gauge an appropriate level of abstraction for Gemini, we conducted an informal study of how people describe animated transition designs.  We recruited five people (3 female, 2 male) near our university campus who had at least 2+ years of data visualization experience. We gave each participant start and end states of an animated transition and asked them to draft an animation design, identifying specific graphic elements, changes to those elements, and timing information. After receiving their first draft, we asked participants to describe possible alternatives. Finally, we showed them implemented animations for the stimuli and asked them to characterize those. All sessions were conducted in person.

To cover multiple transition types and techniques (\eg{} staging and staggering), we chose three stimuli from interactive articles and a video (available in supplemental material). These examples cover 5/7 of Heer and Robertson's transition types~\cite{anim_transition} by omitting \emph{Ordering} and \emph{View Transformation} due to the limited number of recruited subjects. We analyzed participants' transcripts for words indicating timing constraints and graphic components, then derived the following three insights about appropriate abstraction levels:

\begin{enumerate}[label=\textbf{I{\arabic*}.}]
    \item Participants referred to groups of marks by their shape (\eg{} lines, texts, points), role (\eg \q{uncertainty band}), and/or backing data (\q{NY points}). This observation implies that the grammar should be able to select elements by their geometry, roles in the visualization context, or data properties.
    
    \item Participants described changes to guides (axes or legends) both in general terms (\q{expand the y-axis}) and by referring to specific sub-elements (\q{render the x-axis title}).
    
    \item Participants included different staging and staggering elements. Staged animations were described using constraints: synchronizing (\q{at the same time}) and concatenating (\q{then}, \q{after}).
\end{enumerate}

\section{The Gemini Grammar: Motivation and Design}

Gemini is a declarative grammar for specifying animated transitions between two (\emph{start} and \emph{end}) single view visualizations, defined using the Vega grammar~\cite{vega}. By ``single view,'' we refer to charts with at most one x-axis and one y-axis.
The Gemini compiler processes a specified transition design (\autoref{fig:eg1_line}) and the provided start and end states to produce a playable animation plan. 
When designing Gemini, we considered three ways to specify animations: (1) extend a single visualization specification, (2) define transformations that map a start state to an end state, or (3) specify transitions relative to explicit start and end states. We discuss each in turn.

\begin{figure}[!t]
  \includegraphics[width=1.00\columnwidth, trim={0 0 0 0}]{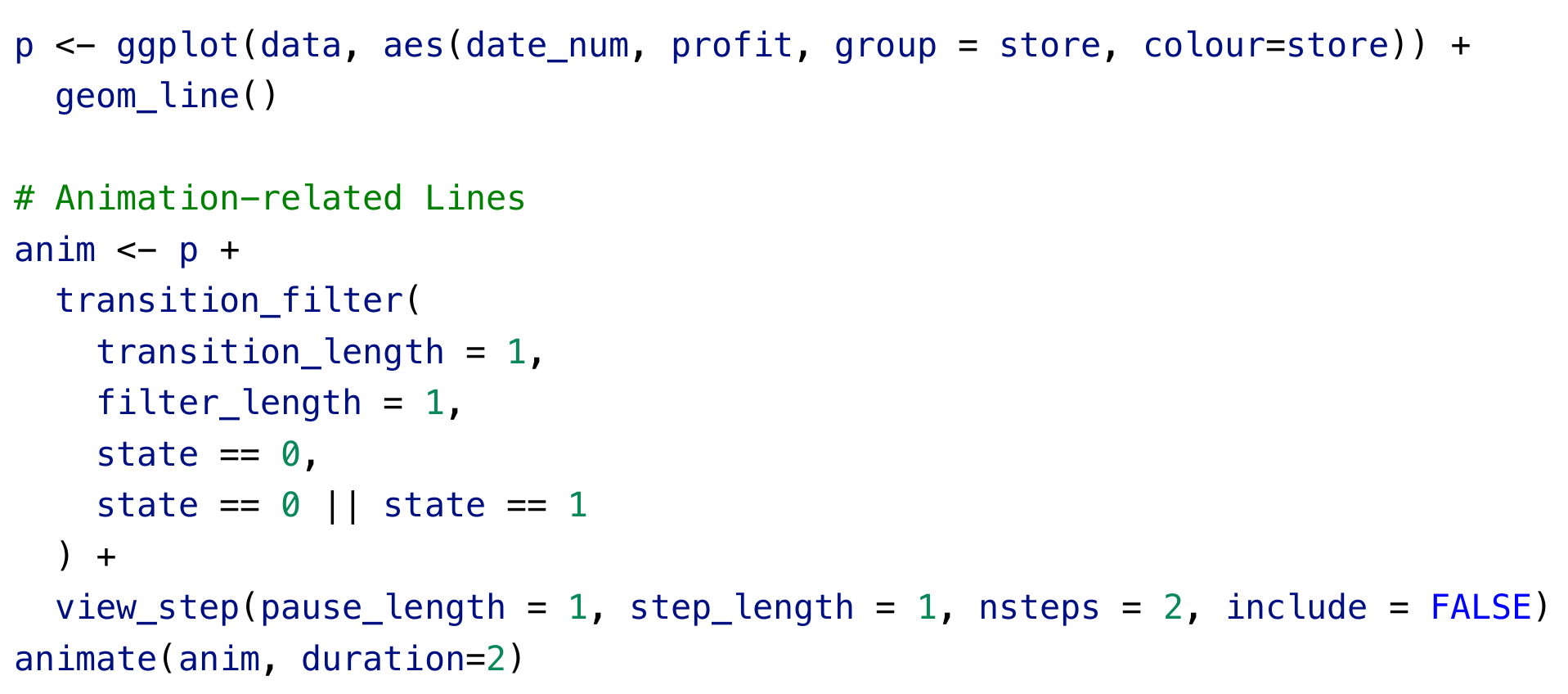}
  \vspace{-18pt}
  \caption{gganimate code to produce \protect\autoref{fig:eg1_line}. Given a ggplot2 chart (\code{p}), gganimate's \code{transition\_filter} and \code{view\_step} are appended to animate the lines between two states and a scale change, respectively.}
  \label{fig:gganimate}
  \vspace{-0.50cm}
\end{figure}

Animation specifications can \emph{extend} existing visualization specifications, as in gganimate~\cite{gganimate}. As \autoref{fig:gganimate} shows, authors can append animation directives to existing ggplot2 code to specify the transition in \autoref{fig:eg1_line}. In this approach, the animation design context closely matches the visualization context and is similar to the animation model of CSS Transitions for Web design~\cite{css}. However, it limits expressible transitions to variations of a single visualization specification, typically data changes under static visual encodings. Since we aim for a broader spectrum of transitions, we do not use this approach.


\begin{figure}[!t]
  \centering
  \includegraphics[width=1.00\columnwidth,  trim={0 0 0 0}]{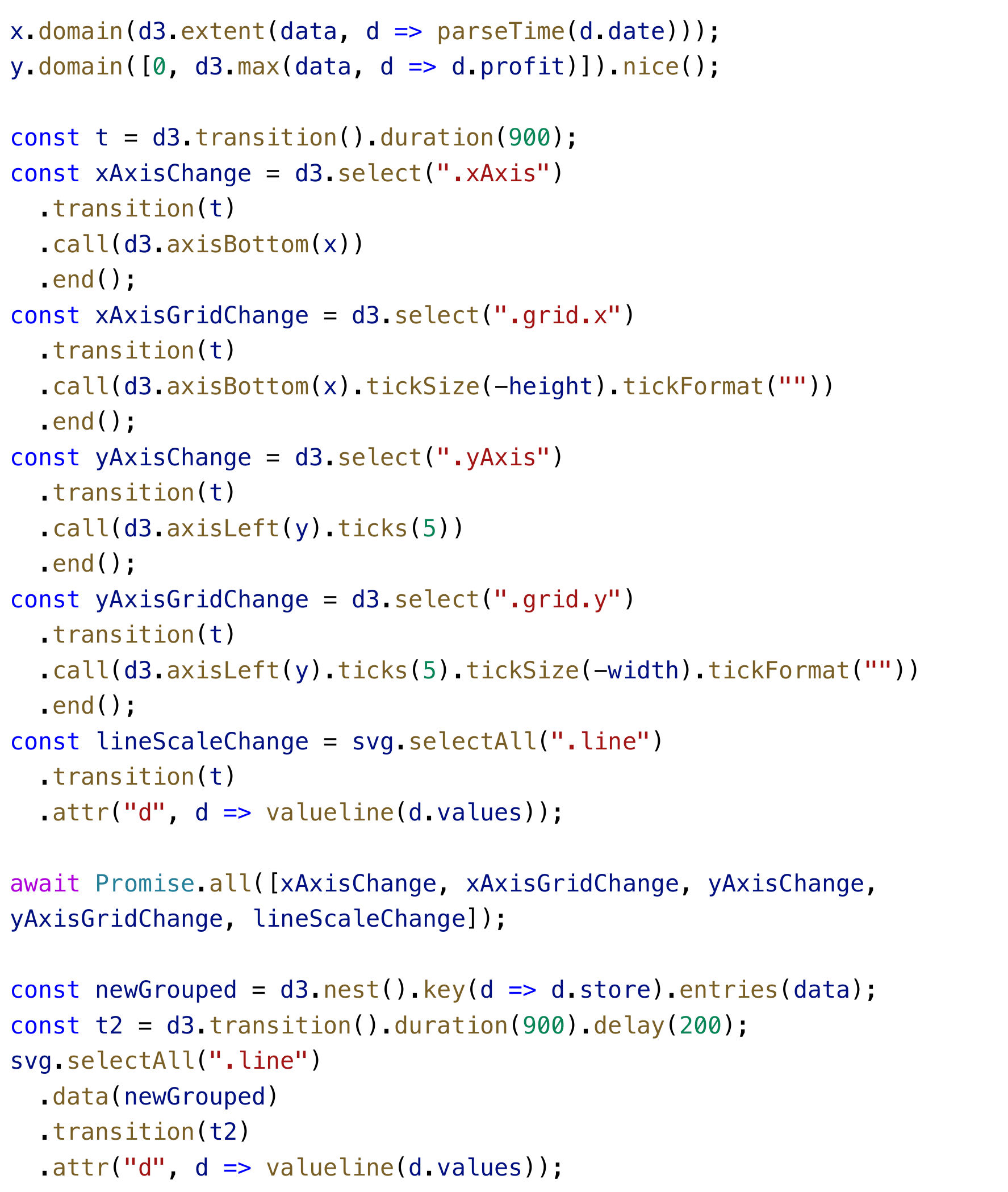}
  \vspace{-18pt}
  \caption{D3 implementation of \protect\autoref{fig:eg1_line}. D3 requires transformations that map the starting visualization state to the end state.}
  \label{fig:d3-example-code}
  \vspace{-0.5cm}
\end{figure}

An alternative is to specify \emph{transformations} that turn a starting visualization state into an end state. Rather than define explicit start and end states separately, \emph{the end state is implicitly defined} by the transformations applied.
For instance, in order to zoom, D3~\cite{d3} users can: create new x and y scales; select the existing axes, grids, and mark elements; and assign the new scales to change the elements to final states (\autoref{fig:d3-example-code}).
This approach requires the definition of all specific manipulations necessary to produce both the animation and end state. It thereby interleaves animation and visualization design: one must redefine an independent target state, in terms of desired transitions.

A third approach is to provide explicit start and end states and then describe desired \emph{transitions} between them. This strategy separates a design into three specifications: two for visualization states and one for the transition. 
As a result, the transition specification can be more succinct, as it needs only to orchestrate the changes to get to the end state \emph{without implicitly specifying the end state}, as in D3's transformation approach. 
However, to tailor animations there must be a way to refer to chart elements across the start and end states, and so a compiler must identify corresponding elements.

Gemini follows this third approach: we aim to support an authoring scenario in which designers (or systems) have explicit start and end visualization states (or ``keyframes''). This approach allows designers to focus on these states before introducing animation concerns, and accords with existing tools. For example, visual analysis tools (\eg{} Tableau, Voyager~\cite{voyager}) already define visualization states using separate, declarative specifications and so are more amenable to this approach. This approach can support a keyframe-authoring paradigm for animating data graphics, which designers reportedly prefer~\cite{anim_authoring_env}.


We now describe the Gemini grammar in detail.
A Gemini specification uses two primary abstractions: the \emph{step}, a basic unit of animation, and the \emph{timeline}, which composes units. 

\subsection{Step: Unit Transitions}

A \emph{step} is a unit transition that interpolates changes to a selected graphic component according to specified timing parameters. The numbered code blocks in \autoref{fig:eg1_line} are examples. Using steps, a complex transition can be split into digestible pieces. Formally, a step is a four-tuple: 
\[
step \coloneqq (component, change, timing, enumerator)
\]

\subsubsection{Step Component}

The step \emph{component} indicates the group of visual elements to be changed. There are \emph{mark}, \emph{axis}, or \emph{legend} components, which can be selected by name. This high-level of abstraction partially satisfies observations from our study, where users select groups of guide elements and marks with the same geometry (\textbf{I1}). Gemini can also control more nuanced components, for example by staggering marks by data fields, or referencing sub-elements of guide elements (\eg ticks, labels, title). In addition, a \emph{view} component refers to the overall frame (\eg chart sizing), and a \emph{pause} (see \codeblockTwo{} in \autoref{fig:eg1_line}) induces delays.


\begin{figure}[!t]
  \centering
  \includegraphics[width=1.00\columnwidth, trim={0cm 0cm 0cm 0cm}]{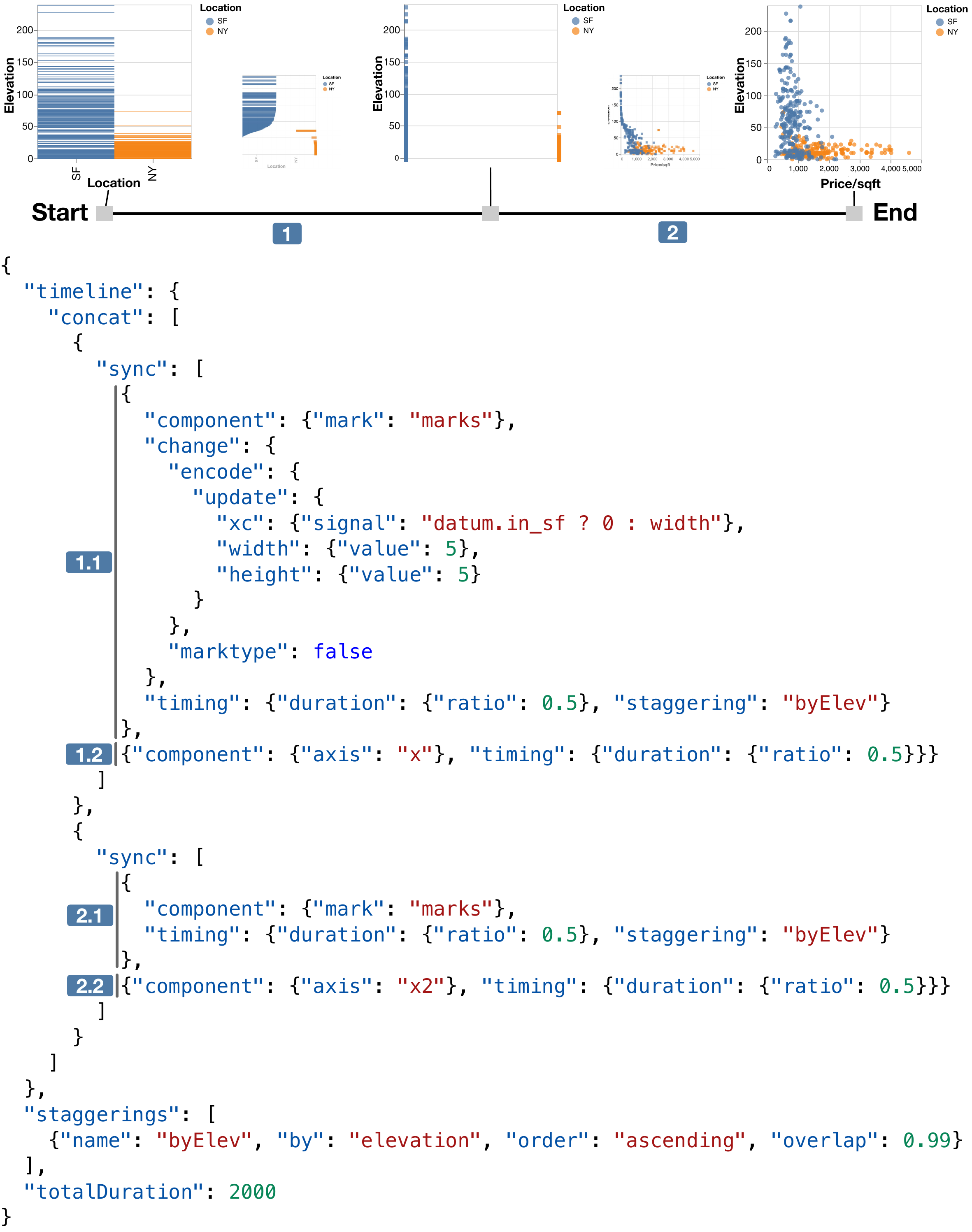}
  \vspace{-15pt}
  \caption{A Gemini specification example for the first part of the animated transition in R2D3 Part 1~\protect\cite{R2D3}. The dot plots (\protect\code{"elevation"} vs \protect\code{"in\_sf"} (=`Location')) are transformed to a scatter plot (\protect\code{"elevation"} vs \protect\code{"price\_per\_sqft"}). \protect\codeblockOneDotOneBlue{} specifies the encoding of the point marks to describe the intermediate state, and the mark steps (\protect\codeblockOneDotOneBlue{} and \protect\codeblockTwoDotOneBlue{}) are staggered to move the points with lower \protect\code{"elevation"} first.}
  \vspace{-15pt}
  \label{fig:r2d3}
\end{figure}


\subsubsection{Step Change}

An animated \emph{change} to a component is defined as a five-tuple:
\[
change \coloneqq (data, encode, scale, signal, mark\mhyphen{}type)
\]
The \emph{data} entry indicates changes to the data backing a mark component, including inserts, removes, updates, and aggregations of the data. For example, \codeblockThree{} in \autoref{fig:eg1_line} performs a data change: it inserts new data from the end state, extending the domain of the line marks. Data changes are applicable only to mark components.

In Vega, datasets are defined by data sources and transformations. Assuming that the start and end states use the same data sources, Gemini identifies data changes by comparing the data transformations. If the dataset is aggregated (or disaggregated), Gemini binds the aggregated values to the raw data so that the raw data can exit (or enter) while being interpolated to (or from) the aggregated values. If no aggregation occurs, Gemini directly joins datasets. If the datasets are grouped by the same data fields, Gemini uses the grouping fields as join keys. Otherwise, it takes user-provided fields or data list indices as defaults. Akin to D3's data join~\cite{d3}, Gemini internally classifies the joined data into an enter set (newly introduced data), update set (persistent data), or exit set (stale data). Joins for these sets can be deferred in order to separate those changes into different animation stages. 

The \emph{encode} entry controls changes to a component's visual encodings. By default, Gemini interpolates between start and end visualizations. For entering and exiting data, the default transition is to fade marks in or out. Encoding changes can be separated into different steps. Further, Gemini lets users directly specify temporary encodings to define intermediate states. For example, \codeblockOneDotOneBlue{} in \autoref{fig:r2d3} shrinks marks toward the sides before settling them into their final positions. For guide components, encode changes can separately target sub-elements to enable fine-grained control (\textbf{I2}). Particularly for the enter set, initial encodings can be set to match the given start state encoding. 

The \emph{scale} entry defines changes to the scales applied to the component. A subset of scales can change between the start and end visualizations. For guide components (\eg axes and legends), scale changes can produce new data for sub-elements, such as axis ticks/gridlines/labels, and legend symbols/labels. Especially for axis components, new data replace old data without joining (i.e., the old ones fade out, and the new ones fade in) when the dimension of the scale's domain changes (e.g., price(\$) to square feet($ft^2$))
By default, Gemini automatically detects the change by comparing backing data fields of the corresponding scales in the visualizations specs. A change of dimension (or lack thereof) can also be explicitly indicated by users.

The \emph{signal} entry refers to changes in Vega signals (dynamic variables) and can be controlled in the same way as \emph{scales}. The \emph{mark-type} entry signifies changes to the geometry of the marks.

By default, Gemini assumes that all changes should be applied concurrently, resulting in a direct interpolation. More elaborate animations can be achieved by specifying more nuanced stages.
For example, the \codeblockOneDotThree{} block in \autoref{fig:eg1_line} suppresses the data change, deferring that change until after the chart scales up (\codeblockThree{}).

\subsubsection{Step Timing}

Every step requires \emph{timing} to schedule changes in a component. This element consists of four properties:
\[
timing \coloneqq (duration, delay, ease, staggering)
\]
The \emph{duration} entry specifies the length of the step, either as an absolute value in milliseconds or as a fraction of the total duration. The \emph{delay} entry is specified similarly and imposes a delay before enacting the change. The \emph{duration} should be provided. If either \emph{duration} or \emph{delay} if expressed as a fraction, the total duration should be specified at the root level. The \emph{ease} property selects a temporal distortion between the progress of the change and the elapsed time~\cite{siso}. Matching D3~\cite{d3}, the default \emph{ease} value is cubic slow-in slow-out pacing.

\begin{figure}[!t]
  \includegraphics[width=1.0\columnwidth]{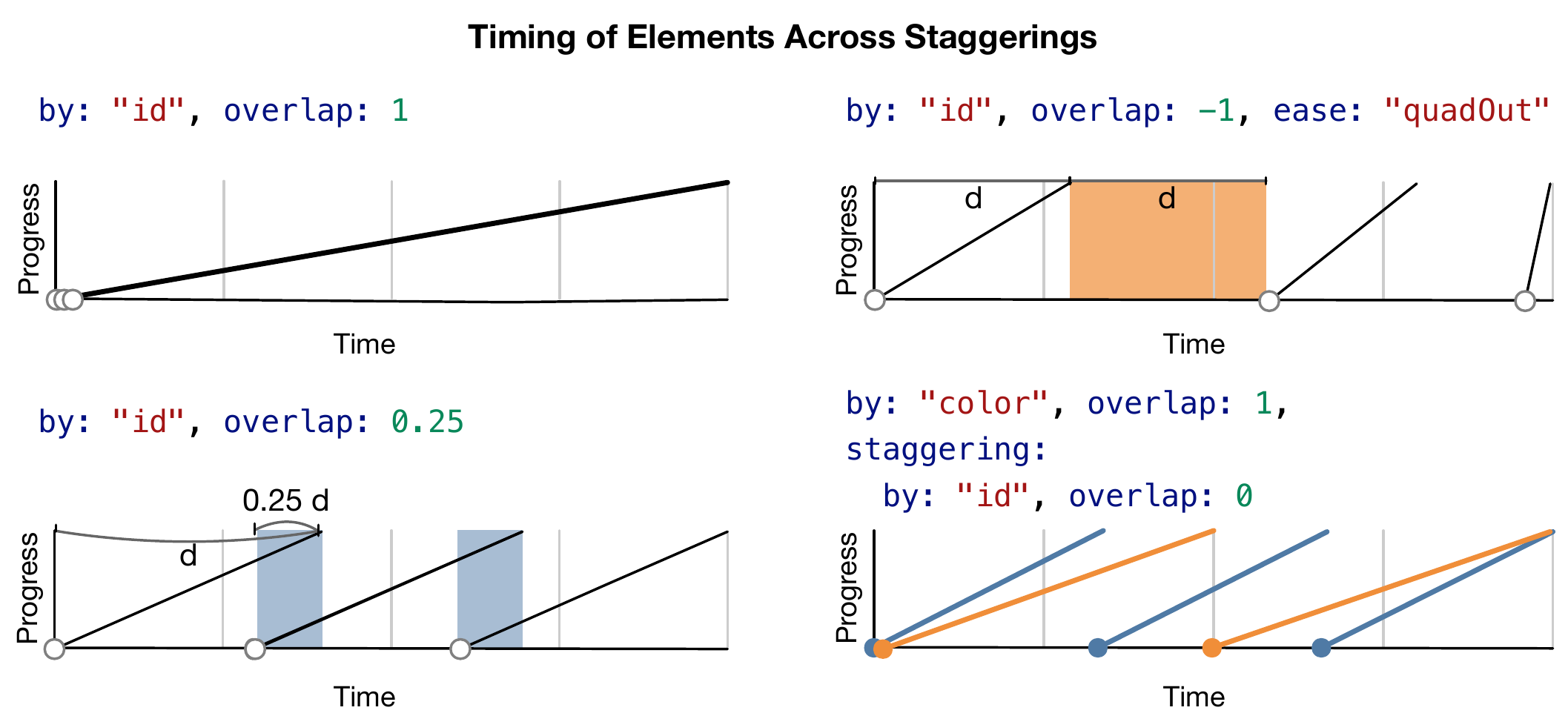}
  \vspace{-20pt}
  \caption{Timing of elements in staggered animations. Points represent individual elements, and lines indicate progress over time. The \code{overlap} parameter controls the intervals between consecutive elements, while \code{ease} distributes the duration of the step across the elements. 
  }
  \label{fig:staggering}
\end{figure}

\emph{Staggering} is applied to step timing by referring to a named staggering specification defined at the root level (\codeblockOneDotOneBlue{}, \codeblockTwoDotOneBlue{} in \autoref{fig:r2d3}). We place staggering definitions at the root level so they can be shared by multiple components. Gemini extends Chevalier et al.'s~\cite{staggering} definition of staggering to include five properties:
\[
staggering \coloneqq (data\,field, order, overlap, ease, staggering)
\]
The \emph{data field} and \emph{order} entries determine grouping and staggering order per element. The \emph{overlap} parameter controls the temporal overlap between the consecutive elements using the ratio:
\[
overlap = \frac{end(elem_i) - start(elem_{i+1})}{end(elem_i) - start(elem_i)}
\]
where $start$ and $end$ are the element's start and end times. The ratio indicates how much an element's change overlaps with its preceding element. The overlap should be less than or equal to 1.0. \autoref{fig:staggering} shows how overlap changes are staggered. The \emph{ease} parameter determines how the duration is distributed to each element. As shown in the top-right of \autoref{fig:staggering}, fast-out easing can be used to assign a smaller amount of time to later elements, emphasizing changes for the first few items and then overlapping subsequent movement. Gemini also supports nested \emph{staggering} for elements in a subgroup (\autoref{fig:staggering}, the bottom-right). 

\subsubsection{Step Enumerator}
\begin{figure}[!t]
  \centering
  \includegraphics[width=1.00\columnwidth]{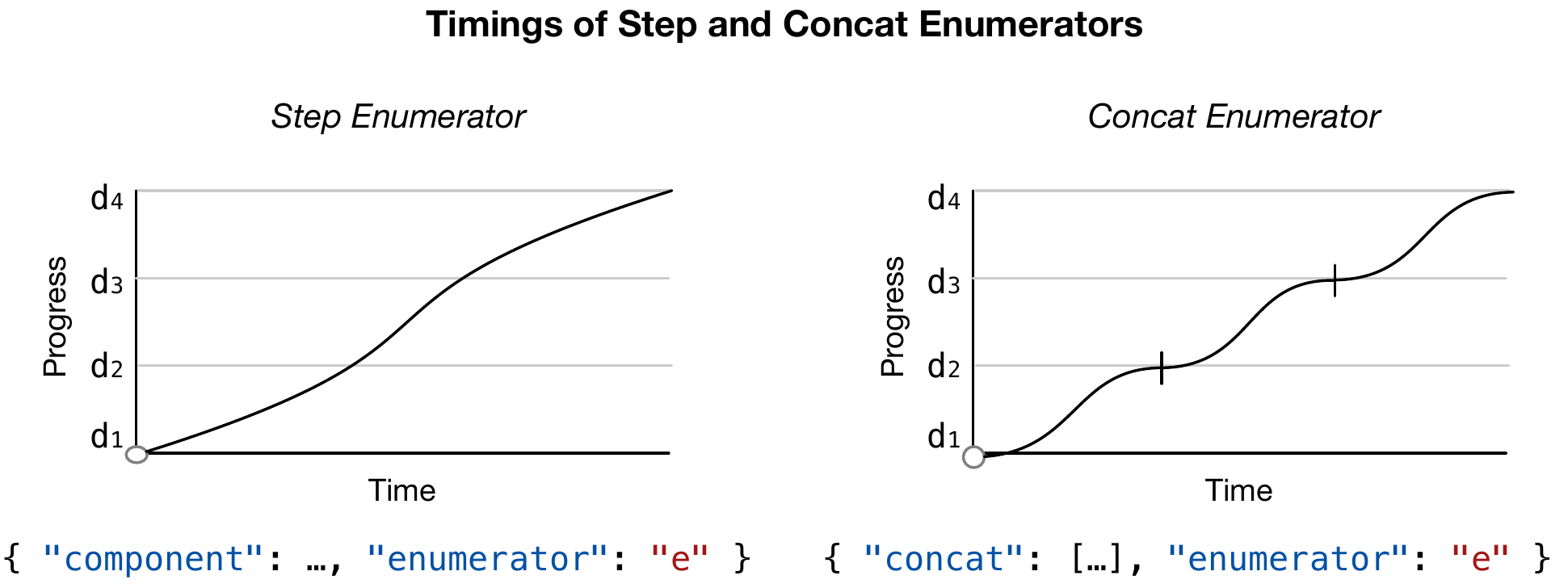}
  \vspace{-18pt}
  \caption{Example timing of an enumerator applied to a step (left) or a concat block (right). The enumerator consecutively joins data sets ($d_1 \rightarrow{} d_2 ...$). Step enumerators iterate the changes within a single step's timing. Concat enumerators iterate changes as multiple steps.}
  \label{fig:enumerator-timing}
  \vspace{-12pt}
\end{figure}

\begin{figure}[!t]
  \centering
  \includegraphics[width=1.0\columnwidth]{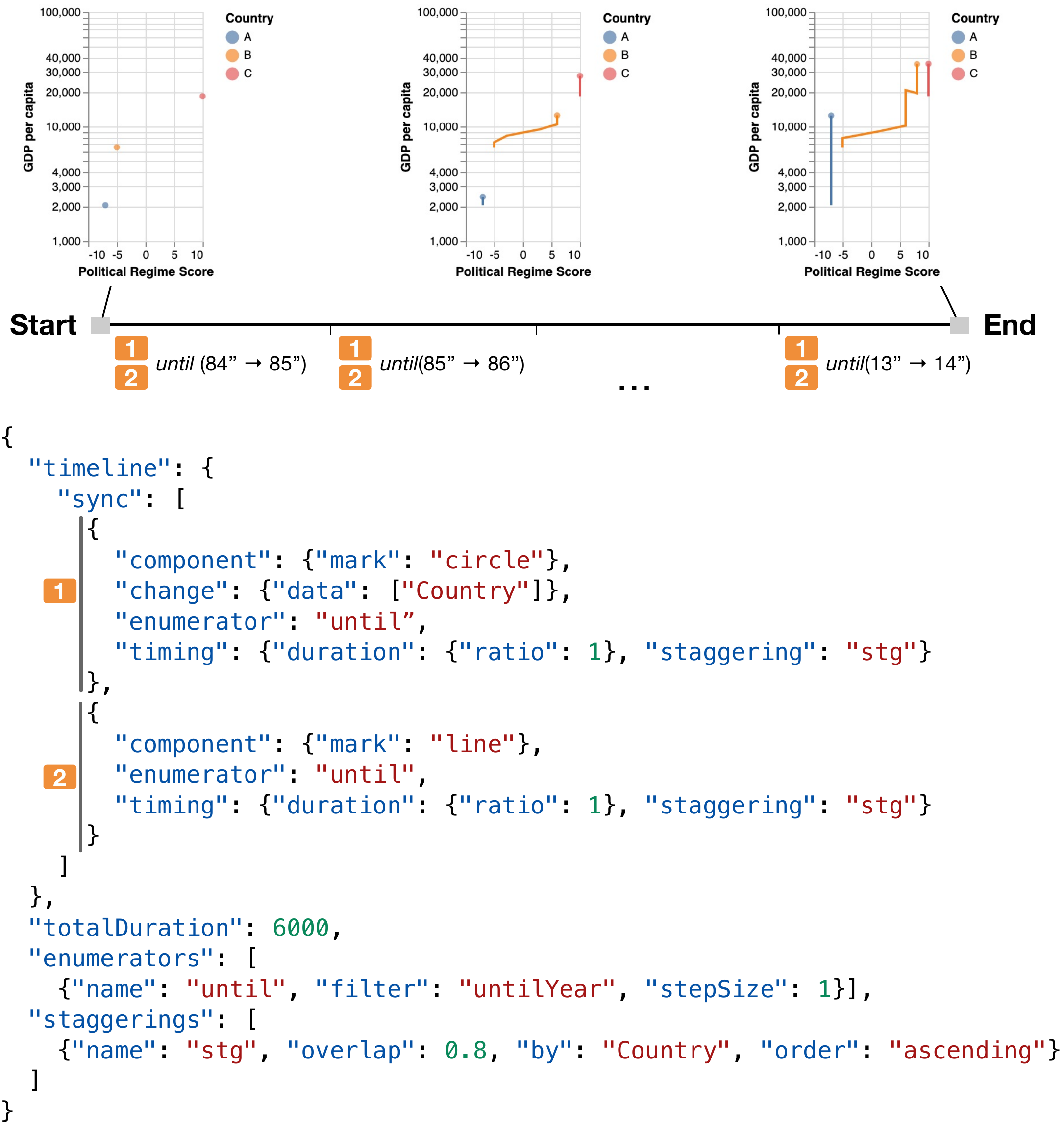}
  \vspace{-0.7cm}
  \caption{Gemini specification for moving points along temporal trajectories (\textit{c.f.},~\cite{nyt_interpreter}). The same enumerator is applied to iterate \protect\codeblockOneOrange{} and \protect\codeblockTwoOrange{}. These steps are staggered so that each point starts and ends individually. }
  \label{fig:enumerator}
  
  \vspace{0.5cm}
  
  \includegraphics[width=1.0\columnwidth]{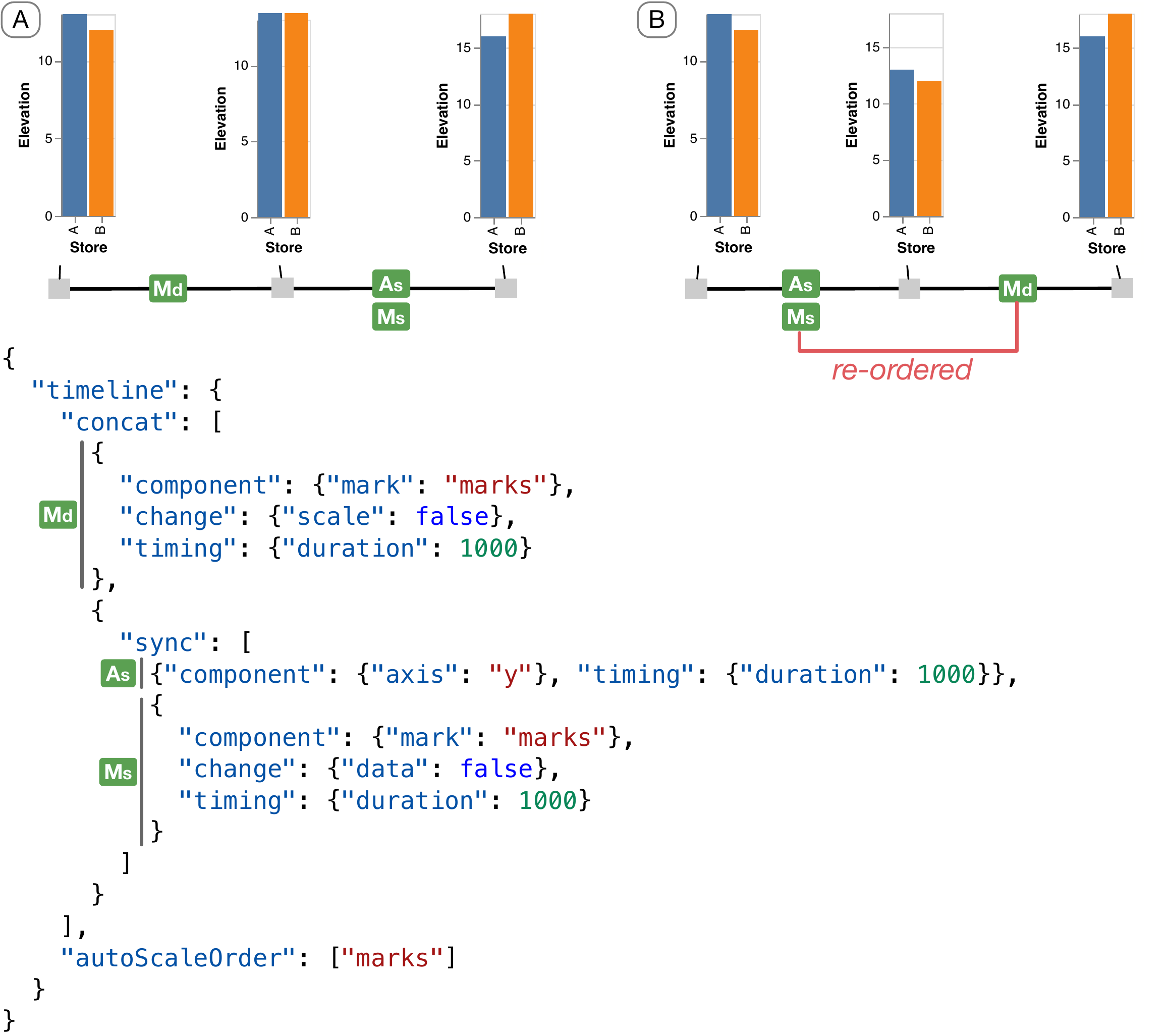}
  \vspace{-0.7cm}
  \caption{Gemini specification for updating the values of bars. The concat block with \protect\emph{autoScaleOrder} automatically swaps the order of its two stages, (\{\protect\codeblockMd{}\} and \{\protect\codeblockAs{},\protect\codeblockMs{}\}), to prevent overflow on the vertical scale.}
  \label{fig:autoScaleOrder}
  \vspace{-0.6cm}
\end{figure}

\[
enumerator \coloneqq (filter, step\mhyphen{}size | [value_1, ...])
\]
An \emph{enumerator} defines a series of data changes (\eg values for specific years) so that it can express consecutive data changes (\eg showing one year at a time). 
When an enumerator is added to a step, Gemini produces a corresponding Vega filter transform and calculates data sets by replacing the filter expression's right-hand side value with enumerated values. These filtering values are derived from the start and end states with a given \emph{step-size}, or provided as an array of values. The calculated data sets are consecutively joined, and these iterations are distributed within a single step (left of \autoref{fig:enumerator-timing}). This approach is Vega-specific, as Vega filter transforms are used to obtain data set selections at a specific moment. \autoref{fig:enumerator} shows an example that uses an enumerator (\codeblockOneOrange{} and \codeblockTwoOrange{}) to depict trajectories of three points over time. 




\begin{figure*}[!t]
  \centering
  \includegraphics[width=2.05\columnwidth,  trim={{0.05\columnwidth} 0 0 0}]{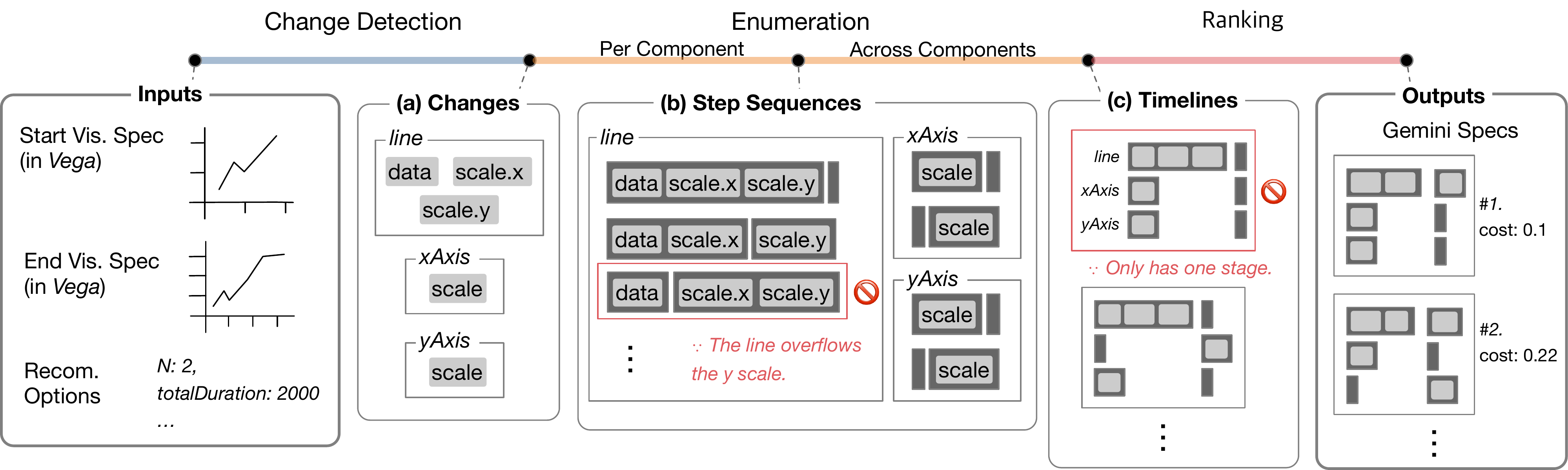}
  \vspace{-20pt}
  \caption{The Gemini recommendation workflow. The system takes as input the start and end visualization states of a transition along with user-provided options, such as the number of animation stages and whether axes maintain their domain dimensions. The system processes the inputs in three steps: (a) it \emph{detects} changes in each component, (b) it \emph{enumerates} candidate timelines by combining detected changes while pruning according to expressiveness constraints, and (c) it evaluates and \emph{ranks} the timelines using its heuristic cost function. }
  \label{fig:recom-overview}
  \vspace{-10pt}
\end{figure*}

\subsection{Timeline: Orchestrating Steps}

Animations can be staged by composing steps into a \emph{timeline}. A timeline entry (or \emph{block}) consists of either a step or one of the composition operations \emph{sync} and \emph{concat}. The sync operation synchronizes an array of blocks at the start or the end based on the \emph{at} property (default is start). The concat operation plays an array of blocks in a sequence. For example, the timeline of \autoref{fig:eg1_line} executes \codeblockOneDotOne{}, \codeblockOneDotTwo{},  \codeblockOneDotThree{} at the same time, but it runs \codeblockTwo{} and \codeblockThree{} step-by-step. Formal representations for block and composition operations are:
\begin{align*}
block &\coloneqq step \, | \, sync \, | \, concat \\
sync &\coloneqq ([block_1, ...], \, at) \\
concat &\coloneqq ([block_1, ...], \, enumerator, \, autoScaleOrder)    
\end{align*}
These operations align with our study observations that interviewees tended to describe animation stages using temporal constraints (\textbf{I3}).

In addition to explicit composition, Gemini provides timeline flexibility via \emph{enumerators} and \emph{autoScaleOrder}. Enumerators for a concat block have the same syntax as step enumerators, although they iterate block-by-block with a divided duration equal to the original duration. \autoref{fig:enumerator-timing} illustrates the difference between these enumeration styles.

The \emph{autoScaleOrder} parameter ensures that the named components' data does not exceed its scale domains by sorting the children blocks of the concat. \autoref{fig:autoScaleOrder} \designA{} and \designB{} show an example that updates quantitative values of a bar chart (\codeblockMd{}), adjusting the scale (\codeblockAs{}, \codeblockMs{}) according to the new values. If the values decrease (\designA{}), the animation updates the values before changing the scale (\codeblockMd{} $\rightarrow{}$ \codeblockAs{}, \codeblockMs{}); otherwise, the old data may overflow the new, smaller scale. If the values increase (\designB{}), the scale changes before the data updates (\codeblockAs{}, \codeblockMs{} $\rightarrow{}$ \codeblockMd{}); otherwise, the new data can overflow the old scale. The autoScaleOrder property lets a single Gemini spec handle these data-dependent cases. 

Users can specify \emph{autoScaleOrder} using the name of a mark component. The Gemini compiler then generates all permutations of concat children blocks that use the specified marks (\eg \codeblockMd{} and the sync of \codeblockAs{}, \codeblockMs{}), and picks one without scale overflow. If the compiler cannot find one, it returns a warning message and uses the original order.

\section{Recommending Animated Transitions}

Leveraging its grammar, Gemini can recommend animation designs by systematically enumerating and ranking transitions. This process can be parameterized using design options, such as the number of transition stages ($N$) and the total duration. Gemini generates suggestions in three steps (\autoref{fig:recom-overview}): change detection, enumeration, and ranking. Our current implementation supports transitions between Vega specifications compiled from single-view Vega-Lite charts. The live version of this implementation is available at \url{https://uwdata.github.io/gemini-editor/}.

\subsection{Change Detection}

Gemini's recommendation system first analyzes differences between input visualizations. It pairs the marks, scales, axes, and legends by their names. Scale domain values, such as numeric ranges and sets of nominal values, are compared to determine whether changes occurred. For changes that Gemini finds hard to detect, users can provide direct input; for example, if the join keys of the data changes or if the domain dimension of a scale changes.

Detected changes are grouped per component, as in \autoref{fig:recom-overview}~(a). Each change is one of the change types supported by the Gemini grammar (data, scale, encode, mark-type, signal) and contains meta information, such as initial and final states of the corresponding component, and whether the component expands or shrinks the width and height of the chart. Gemini separates the encode changes of mark components into high-level channels (x, y, color, shape, size, opacity, text) so that generated animations can stage these encode changes separately.

\begin{table}[!t]
  \caption{Constraints for pruning enumerated step sequences. Sequences violating any of the constraints are excluded.}
  \label{tab:exp-constraint}
  \scriptsize
	\centering
  \begin{tabu}{ p{1.2cm} p{6.8cm} }
  \toprule
   \textbf{Constraint} & \textbf{Description}\\
  \midrule
   Unavailable Scale & Any scale that is not available, e.g., changing a point mark's encoding to have color using a color scale but without the scale change introducing the color scale.\\
  \midrule 
   Unavailable Data Field & Any data field that is not available, e.g., changing data to aggregate without changing the encoding so the prior encoding refers to an old data field not in the newly aggregated data.\\
   \midrule
   Unavailable Encoding & Any visual attribute that is not supported by its mark-type, e.g., changing the mark-type from point to rectangle without changing the encoding, such that the rectangle mark does not have encodings for width or height.\\
   \midrule
   Overflow & Data overflows any domain of the scales used in the encoding, e.g., changing a bar chart's data to include 3 more categories without changing the corresponding categorical scale so newly introduced categories cannot be represented.\\
  \bottomrule
  \end{tabu}%
  \vspace{-0.5cm}
\end{table}

\subsection{Timeline Enumeration}

Gemini enumerates candidate timelines by combining detected changes. The enumeration is processed in two steps. First, for each component, Gemini enumerates sequences consisting of $N$ sets of changes, where $N$ is the target number of animation stages. Gemini prunes sequences inducing illegal intermediate encodings or data overflow, as shown in \autoref{tab:exp-constraint}. For example, consider a transition that expands a line chart so that data and scales of the marks (lines) change as shown in \autoref{fig:recom-overview} (or \autoref{fig:eg1_line}). The mark component (\emph{line}) has three changes, \code{scale.x}, \code{scale.y}, and \code{data}. These three changes can be combined into a two-stage animation in 8 ($=2^3$) ways. Among the possible sequences, Gemini filters out the three that cause overflow ([\{\code{data}\}, \{\code{scale.x}, \code{scale.y}\}], [\{\code{data}, \code{scale.x}\}, \{\code{scale.y}\}], [\{\code{data}, \code{scale.y}\}, \{\code{scale.x}\}]) by expanding the data before expanding both scales. The sets in each resulting sequence correspond to steps in the Gemini grammar.

Gemini, in turn, iteratively picks one of the resulting sequences per component and combines selections, while excluding combinations having an empty stage. For instance, if Gemini enumerates two components ($A, B$) involving one change for 2-stage designs ($change(A) = \{\code{a}\}$, $change(B) = \{\code{b}\}$, $N=2$), it first enumerates step sequences ($seq(A)$ = \{ [\{\code{a}\}, $\varnothing{}$], [$\varnothing{}$, \{\code{a}\}] \}, $seq(B)$ = \{ [\{\code{b}\}, $\varnothing{}$], [$\varnothing{}$, \{\code{b}\}] \}). Then Gemini enumerates complete sequences ([\{\code{a}\}, \{\code{b}\}], [\{\code{b}\}, \{\code{a}\}]) while excluding those with an empty stage ([\{\code{a,b}\}, $\varnothing{}$], [$\varnothing{}$, \{\code{a,b}\}]).
Gemini then reviews the combined sequences and inserts view changes that expand and shrink the width or height of the chart. An expanding view step is inserted in the first stage that expands the chart, and a shrinking view step is inserted on the last stage that shrinks the chart. 
Gemini evenly distributes the user-specified total duration across the stages and leaves further customization to users.

\subsection{Transition Ranking}

Gemini evaluates all enumerated timelines using a scoring function that attempts to quantify the complexity of a Gemini animation specification. We define ``complexity'' as a proxy measure of how much the interpretation cost exceeds an assumed capacity: ``cost'' models the effort viewers must expend to identify animation changes, while ``capacity'' refers to how much cost viewers can tolerate. Gemini uses a heuristic function to rank enumerated timelines in ascending order of complexity (lower is better). We make the following assumptions: 
\begin{enumerate}
    \item \emph{Cost and capacity}: Each change type has a positive cost, and people have a positive capacity per stage. 
    \item \emph{Capacity as a function of duration}: Capacity monotonically increases with the duration of a stage.
    \item \emph{Bundling effects}: Some animated changes may become easier or harder to perceive if they are synchronized.
\end{enumerate}

\begin{table}[!t]
  \caption{Bundling effect conditions ($B$). The first three conditions maintain valid data graphics during the transition~\cite{anim_transition}; the second one applies the Gestalt common fate effect. The last two conditions encourage bundling of spatial or non-spatial changes.}
  \label{tab:bundling}
  \scriptsize
  	\centering
  \begin{tabu}{ p{7.2cm}  p{0.8cm} }
  \toprule
   \textbf{Condition} & \textbf{Effect} \\
  \midrule
   Apply scale changes with different domain dimensions while not changing the corresponding encodings, e.g., new ``temperature'' scale applied on the old encoding to ``precipitation'' data field. & Penalty\\ 
   \midrule
   Change the x(/y) scale of the mark and x(/y)-axis together. & Discount\\
   \midrule
   Change the non-spatial scale (color, size, shape, opacity) of the mark and the corresponding legend together.  & Discount\\
   \midrule
   Change x and y scales of marks together but do not change their domain dimensions.  & Discount \\
   \midrule
   Change non-spatial scales of marks together. & Discount\\
  \bottomrule
  \end{tabu}%
  \vspace{-10pt}
\end{table}

The first assumption aligns with prior work that finds that too many changes at once are hard to follow~\cite{anim_transition,cap}. We model capacity as a monotonically increasing function of animation duration by hypothesizing that people can identify more changes as they can track slower objects with more time to process them. Finally, we assume bundling effects, which penalize or discount the cost of specific bundles of synchronized changes. For example, if the scale of the y-axis and the mark component’s y-scale change together, viewers may perceive this as one bundled change (``vertical change'') by Gestalt common fate. This approach may be more effective than presenting them separately without synchronization. 
Specifically, the complexity function is:
\begin{gather*}
Complexity = \sum_{s\,\in\,Stages}{max(0,\, W(s) - C(duration(s)) + B(s))} \\
\text{where $W(s) = \sum_{x\,\in\,change(s)}{w(x)}$}
\end{gather*}
$W$, $C$, and $B$ represent the total weighted step cost, capacity, and bundling effect of a stage, respectively. We assign step costs $W$ to follow the edit operation costs of GraphScape~\cite{graphscape}. For example, $w(\text{\emph{mark-type}}) < w(data)$ means that edit operations in the Mark category are cheaper than Add/Remove/Modify Filter and Aggregate operations in the Transform category. We determined the bundling effects in \autoref{tab:bundling} in reference to existing design principles~\cite{anim_transition}. The precise values of the costs and the bundling effects are available as supplemental material. Lastly, we model capacity as a sigmoid function whose output converges to a ceiling as its input increases. We initially fit the parameter of the sigmoid to reflect the scale of the costs and our empirical observations: $C(t) = \frac{0.8}{1+exp(-(t-800)/300)}+0.2$, where t is in ms. These parameters can be adjusted to align with other design guidelines or preferences. 

\subsubsection{Formative Crowdsourced Study}

\begin{figure}[!t]
  \centering
  \includegraphics[width=1.0\columnwidth,  trim={{0.0\columnwidth} 0 0 0}]{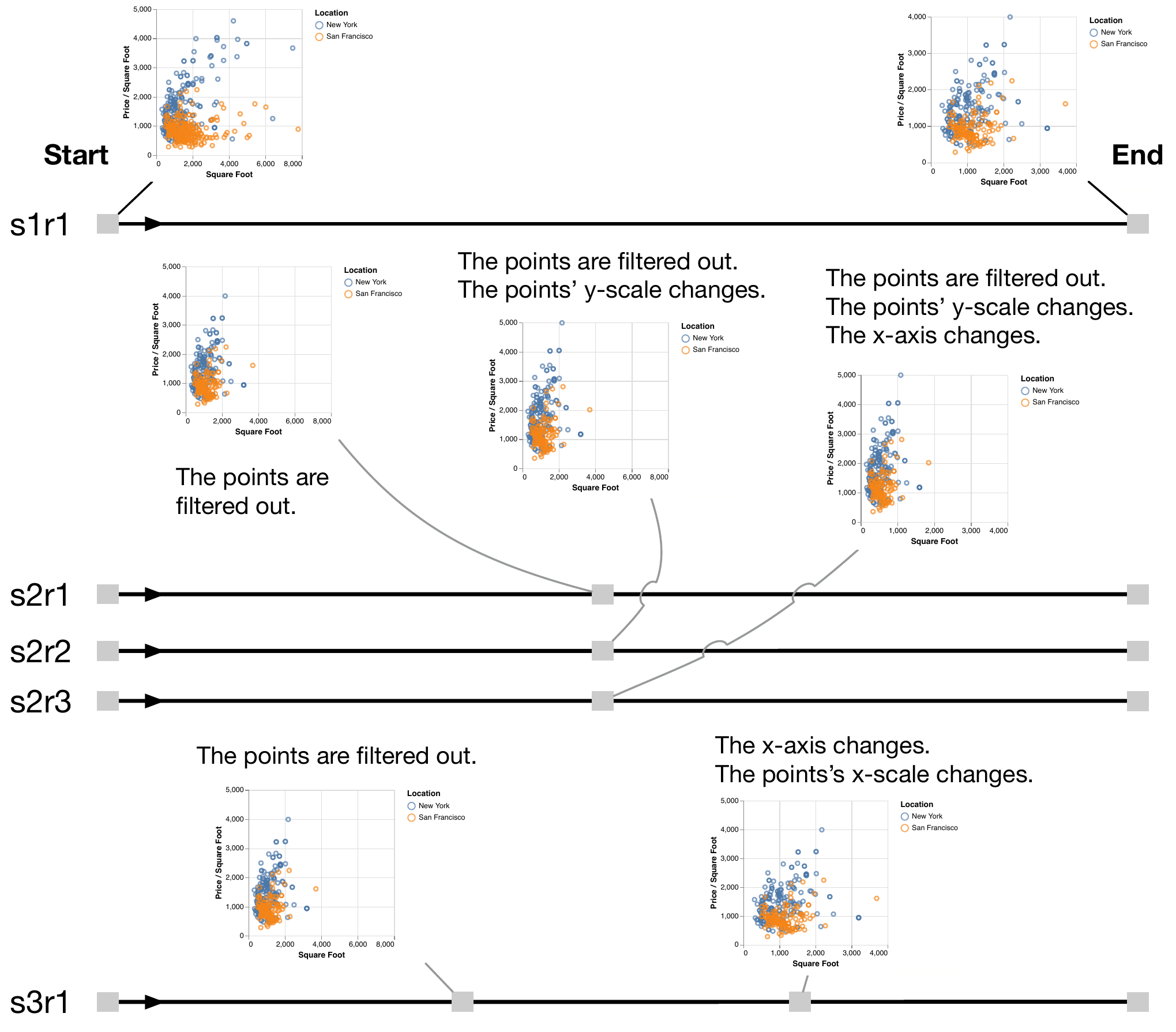}
  \vspace{-20pt}
  \caption{Gemini-generated animation designs for the Filtering Points stimulus of the formative study. The scatter plots represent house price data in two locations, and the transition filters out houses with 3+ beds. The other stimuli are available in the supplemental material.}
  \label{fig:eg-stimulus1}
  \vspace{-10pt}
\end{figure}

To test and refine our recommendation logic, we conducted an online experiment. We recruited 53 people (26 females, 26 males, 1 declined to state) from Amazon Mechanical Turk and asked them to rank animated transitions recommended by Gemini in terms of how well they represent the transition in a clear and logical way. Then, we compare the collected users' rank to Gemini's cost function.

We gave 5 Gemini-recommended animation designs to each subject for each of 4 transitions covering 6 of the 7 transition categories of Heer \& Robertson's transition taxonomy~\cite{anim_transition}. Of the 7, \emph{View Transformation}, which moves camera positions, is omitted since it is unusual in statistical graphics, and scale changes can depict similar effects (e.g., zooming and panning). We picked the following 5 designs per transition: the single-stage design (s1r1), the best 2-stage design (s2r1), the median-ranked 2-stage design (s2r2), the worst-ranked 2-stage design (s2r3), and the best 3-stage design (s3r1). All animations were 2 seconds long, and this duration was evenly distributed to the stages. As an example, one stimulus is shown in \autoref{fig:eg-stimulus1}. The participants ranked the 5 designs of each stimulus at a time. The order of the stimuli and the order of the designs within a stimulus were randomized. Participants were required to play each animation at least twice before ranking them. We also asked subjects to describe what features of the best and worst designs informed their ranking. Participants were compensated \$2.50 USD.

Our analysis excludes responses of participants who submitted irrelevant rationale text (e.g., ``Pixar is best known for CGI-animated...''). To determine significant differences among the designs in each stimulus, we use the Friedman rank-sum test with post hoc pairwise comparisons. We report a user-preference A over B as $A>B$ when the pair has a significant difference ($p<0.05$).

\begin{figure}[!t]
  \centering
  \includegraphics[width=1.00\columnwidth,  trim={{0.00\columnwidth} 0 0 {0.00\columnwidth}}]{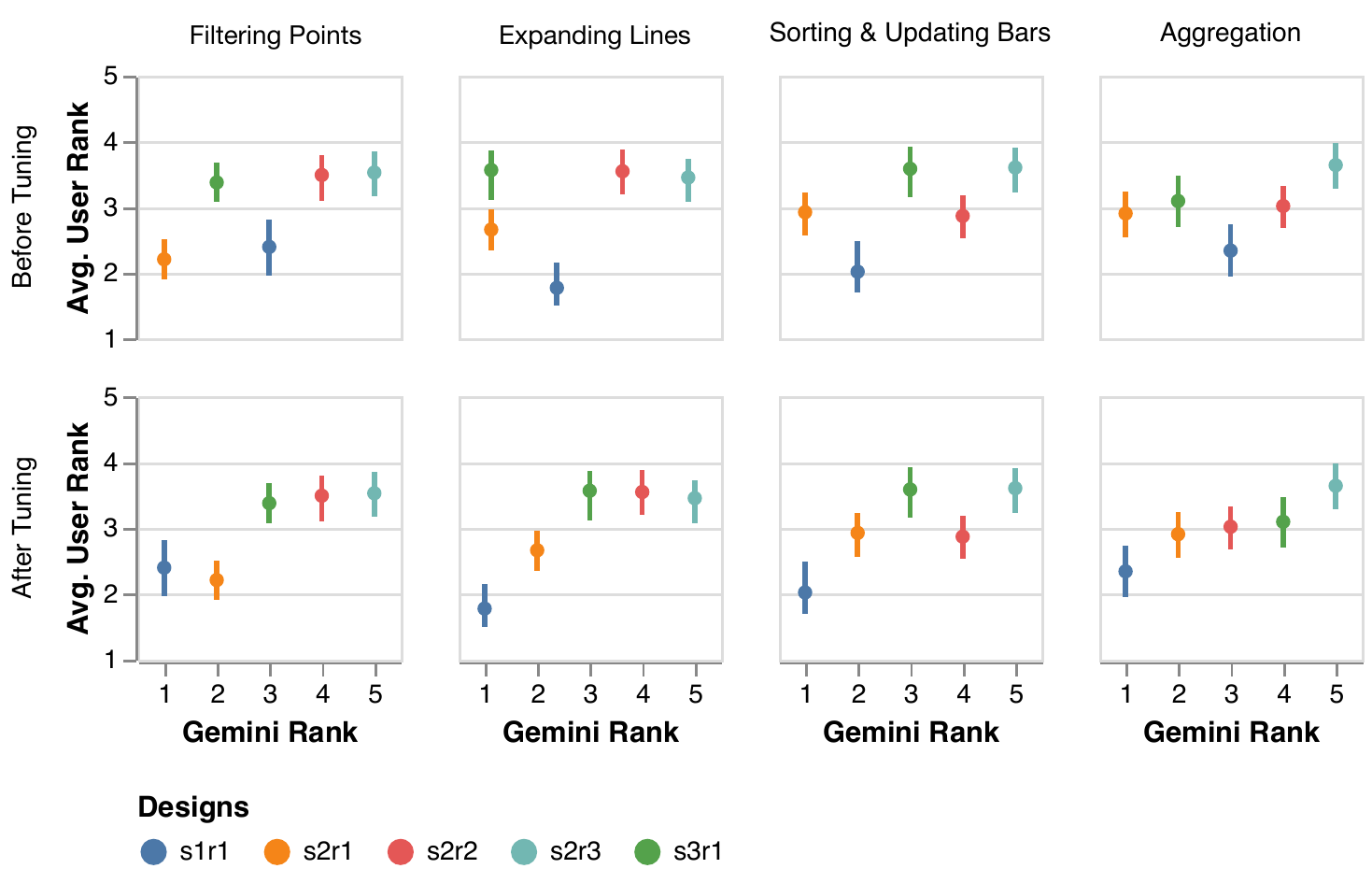}
  \vspace{-20pt}
  \caption{The Gemini rank against the average ranks by the subjects across the stimuli. The line shows the bootstrapped 95\% confidence interval of the ranks. We tuned Gemini costs to better align with the observed rankings by promoting the single-stage design (s1r1) and demoting the 3-stage design (s3r1). We note that s2r1 and s3r1 for Expanding Lines were tied for the top Gemini rank prior to tuning.}
  \label{fig:rank-comparison}
  \vspace{-10pt}
\end{figure}

The charts on the top row of \autoref{fig:rank-comparison} show the experimental Gemini-evaluated ranks against the average of the subjects' ranks per stimulus. Overall, Gemini underestimates user preferences for single-stage transitions (s1r1 $>$ *), while over-estimating preferences for 3-stage transitions (s3r1 $<$ s1r1, s2r1). Participants most favored s1r1 designs since they are smooth and slow. Some preferred s3r1 designs the least because they show too many animation effects.

The first stimulus (Filtering Points) filters out the data points and adjusts the x and y scales to the remaining data, exemplifying \emph{Filter} \& \emph{Substrate Transformation} transition types~\cite{anim_transition}. Similarly, the second stimulus (Expanding Lines) expands lines along the time axis while expanding the domain of the two scales (\emph{Timestamp} \& \emph{Substrate Transformation}). In both stimuli, the animations staging the horizontal and vertical scale changes were less preferred than the ones synchronizing them in one stage (s1r1, s2r1 $>$ s2r2, s2r3, s3r1, except s2r1 $>$ s2r3 in Expanding Lines). This observation aligns with Gemini bundling effects that discount the cost when the x and y scales change together (the fourth item in \autoref{tab:bundling}).
 
The third stimulus (Sorting \& Updating Bars) sorts vertical bars and updates their values by substituting the data field (\emph{Order} \& \emph{Schema Change}). Only the s1r1 animation is preferred over other animations (s1r1 $>$ *). Among the other less-favored four, s3r1 and s2r3 are the least favored (though not significantly so) because they sort the bars and x-axis labels separately. This preference corresponds with a Gemini bundling effect that discounts the cost when the axis and mark change the same scale (the second item in \autoref{tab:bundling}).

The last stimulus (Aggregation) aggregates the averages of the data points per group and transforms them into bars (\emph{Visualization Change}). For this transition, participant preferences diverged without significant differences except for s1r1 $>$ s2r3. Notably for s2r3, subjects reported that the rising bars without the corresponding y-axis change at the last stage seemed manipulated. 

Based on the surveyed ranking, we manually adjusted the parameters of the capacity function by decreasing the intercept, increasing the ceiling value, and shifting along the x-axis: $C(t) = \frac{1.4}{1+exp(-(t-1200)/300)}$. By doing so, the curve starts from the lower value and rises up to the higher ceiling value. It increases the capacity of the stages with $t>\; \sim1,373$ while reducing it for the other shorter stages. As a result, Gemini recommendations promote the single-stage animation design (2000ms per stage) and demote the 3-stage designs (667ms per stage). The tuned ranks more strongly correlate (though not perfectly) with the observed user ranks, as shown in the bottom row of \autoref{fig:rank-comparison}. Further systematic and data-driven tuning is a promising future work area relative to our manual adjustment.

\begin{table*}[!t]
  \caption{Replicating user-created D3 animations in Gemini. Subjects spent 40--150 minutes to animate the given transitions in D3. Of 11 animations, 3 exactly match the top-3 Gemini recommendations, and 5 can be derived from the recommendations by adjusting timing parameters only.}
  \label{tab:summary-of-d3-anim}
  \scriptsize
  \centering
  \begin{tabu}{ p{1cm} c c p{4cm} c p{5cm} }
  \toprule
   Stimulus & Subject-Design & Completion Time (min) & Identified Mistakes & Replicable & Required Edits \newline (except for duration \& delay changes)\\
  \midrule
   \multirow{2}{1cm}{Filtering Points} & P1-1 & 60 & The marks instantly disappear. & Yes & \textit{None.} \vspace{5pt} \\
    & P2-1 & 40 & The symbols of the legend moves. & Yes & Add staggering to the first mark step.\\
   \midrule
   
   \vspace{-10pt}\multirow{4}{1cm}{Expanding Lines} & P3-1 & 60 & The lines shrink with some delay (200ms) comparing to the axes. \vspace{5pt} & Yes & \textit{None.} \\
    & P4-1 & 45 & None. & Yes & \textit{None.} \vspace{5pt} \\
    & P4-2 & +15 from P4-1 & None. & Yes & Swap x- and y-axis steps, edit the first mark step to synchronize x-axis and data updates.   \vspace{5pt}\\
    & P4-3 & +30 from P4-2 & None. & No & Not applicable. \\
   \midrule
   
   \vspace{-5pt}\multirow{2}{1.2cm}{Sorting \& Updating Bars} & P5-1 & 150 & None. & Partially & Insert a stage to shrink the bars and the y-axis before updating the data.\vspace{5pt} \\
    & P6-1 & 130 & Axis changes instantly at the end. & Yes & Add staggering to mark and x-axis.\\
   \midrule
    
   \multirow{2}{1cm}{Aggregating} & P7-1 & 55 & The y-axis title does not change. & Yes & Insert a 0 ms stage introducing the bars at aggregated value positions. Change the first mark step to keep prior x-scale. \vspace{5pt}\\
    & P8-1 \vspace{5pt} & 79 & \multirow{2}{4cm}{The y axis does not change, the y scale of the marks does not change, and the width of bars get halved.} & Yes & \multirow{2}{5cm}{Significant edits are required, including appending extra marks to the start and end visualization specs.}\\
    & P8-2  \vspace{5pt} & +22 from P8-1 &  & Yes & \\
  \bottomrule
  \end{tabu}%
  \vspace{-15pt}
\end{table*}

\section{Evaluation: Replicating User-created D3 Animated Transitions Using Gemini}

We conducted a user study in which we collected manually authored animated transitions by designers using D3, and attempted to replicate them using Gemini recommendations. We chose D3 as it is a popular and expressive tool for creating animated visualizations.

Although evaluation studies typically compare user experiences with a proposed tool and other baseline tools, we instead take an experimenter-replication approach for multiple reasons: 1) the baseline tool (D3) has unequal amount of resources (e.g., API documentation, tutorial, examples) for users, and 2) our primary purpose at this juncture is to assess whether Gemini can express and recommend what users want to design, rather than whether the Gemini grammar improves the user experience. 
Accordingly, we do not make any direct claims regarding user experience.
We can, however, verify expressiveness by assessing whether the replications express authors' original designs, and the recommendation quality is implied by the amount of required edits to the recommendations to replicate participants' designs.


\subsection{Study Design}

We recruited 8 (4 female, 4 male) D3 users with data visualization design experience.
Seven participants were graduate students studying Human-Computer Interaction (P8) or data visualization. One (P3) worked at an IT company in a data visualization-related role. Participants used D3 for 2+ years ($\mu=4.5$), though they reported that they do not currently use D3 daily. P3 participated remotely, while the others spent at least 45 minutes in the presence of the first author. After the session, subjects were allowed to complete the task remotely. All subjects waived compensation for the study except for P8, whom we compensated with a \$15 USD gift card.

We gave each participant D3 code for the start and end visualizations for one of the four transitions used in our formative (tuning) study. We asked them to design and implement 1+ animation of the given transitions in D3. They used their own laptops and were allowed to access any resources for the task (e.g., Internet search). After finishing the task, the subjects sent the first author their code with text reports about their prior D3 experience and completion time for each animation.

After collecting all submissions, we noted suspicious aspects that may not have been what the participants intended. These parts were identified by 1) differences between the final states of the subjects' animations and the given end visualization, and 2) static changes without any animation effect. We reached out to each participant to confirm if the suspicious parts were indeed mistakes. After confirmation, we replicated the submissions without mistakes by editing one of the top 3 Gemini recommendations. Finally, we confirmed with participants that our replications matched their original intent.

\subsection{Results}

\autoref{tab:summary-of-d3-anim} summarizes the results. Participants crafted a total of 11 animations for the given transitions. The self-reported average completion time for their first animation design was 77 minutes ($t \in [40, 150]$). The manually authored D3 animations, the top 3 Gemini recommendations, and the Gemini replications derived from the recommendations are available in the supplemental material and \url{https://uwdata.github.io/gemini-d3-study/}.
Based on the results, we now assess the expressiveness of the Gemini grammar and the utility of Gemini suggestions. 

\subsubsection{Expressiveness}
Using Gemini, we successfully replicated 9 of 11 (82\%) participant designs, as corroborated by the participants. Regarding the two replication failures, P5-1 was partially created, but P4-3 was not. P5-1 vertically squeezes and stretches the bars and the y-axis toward the bottom line by SVG scale transformation while fading in and out. Since the Gemini grammar cannot apply this geometric scaling, we achieved the vertical scaling by changing the position encoding of all components except for the y-axis title, which just faded out and in without moving. P5 confirmed that our replication was similar to the original intent except for the title. P4-3 meticulously stages the horizontal and vertical expansions of each component so that the y-scale expands when the expanding lines reach the ceiling. This staged animation requires low-level individual timing of x- and y-scales of the line, which is not supported by the Gemini grammar due to its high-level abstraction.

\subsubsection{Gemini Recommendations as Starting Points}
We found that most authored animations can be derived from Gemini recommendations without significant modifications. Among the 10 expressible user-created animations, 5 can be replicated from one of the top 3 Gemini recommendations by adjusting timing only, such as the duration and delay of steps (P1-1, P3-1, P4-1) or staggering (P2-1, P6-1). Moreover, P4-2, P5-1, and P7-1 can be achieved by editing the recommendations by modifying the existing steps with minor edits (P4-2) or inserting a new stage with one or two steps to elaborate the recommended flow (P5-1, P7-1). Interestingly, P4-2 matches a design rejected by Gemini's recommender due to overflow. In other words, Gemini could recommend this design if we relaxed our expressiveness constraints (\autoref{tab:exp-constraint}). On the other hand, P8-1 and P8-2 require significant effort to express in Gemini, requiring edits to the given visualization specifications to append extra marks and data transformations (P8-1, P8-2) and a parallel staging structure to overlap the animations of different components (P8-2).

\subsubsection{User Performance in Crafting Animation with D3}
Collecting the user-crafted animations and self-reports reveals the effort participants spent on their designs. First, all participants report spending 40+ minutes ($\mu=77$) to arrive at an initial design, which includes the time for running and understanding the given D3 code, ideation, and implementation. Moreover, 7 of 11 designs exhibited mistakes. The mistakes in P1-1, P2-1, P3-1, and P7-1 were made because the authors overlooked differences, and the ones in P6-1, P8-1, and P8-2 were due to time constraints (i.e., the authors did not want to spend additional time.). All mistakes are avoided by Gemini's recommendations, which are generated in under 1 minute on a laptop computer.

\subsubsection{Implications}
These findings imply that Gemini provides reasonable expressiveness and reliable recommendations with which designers can start authoring and exploring animation designs. For example, by employing Gemini, animated transition authoring tools could provide recommendations to facilitate the authoring process with initial designs. 

\vspace{-0.1cm}
\section{Future Work and Conclusion}

We presented Gemini, a declarative grammar and recommender system for authoring animated transitions between single-view statistical graphics. The high-level declarative format of the Gemini grammar provides a representation with which software can generate and explore animated transition designs. To evaluate Gemini, we replicated user-created animated transitions by modifying one of the top Gemini recommendations. Through these replications, we found that Gemini can provide reliable starting points for the authoring process. The Gemini compiler, recommender system, and examples are available as open-source software at \url{https://github.com/uwdata/gemini}. 

There remain challenges for effectively using Gemini.
Our evaluation results imply that Gemini suggestions can serve as reliable starting points for user-desired animation designs.
However, an outstanding issue is the user interface for specifying visualization states and editing Gemini specifications.
While existing APIs~\cite{vega-lite, dziban} and graphical tools~\cite{lyra, voyager, voyager2, polaris} support authoring of visualization states, new interfaces are needed for animation specification.
Future interfaces can lower the threshold for interacting with Gemini specifications (e.g., how graphical tools like Lyra~\cite{lyra} assist creating Vega visualizations) while enabling generation and review of recommendations. 

In addition to user interface concerns, richer recommendations could facilitate design exploration. 
The current implementation of the recommendation system is a proof-of-concept that is limited to transitions between single view charts without layers. Additional types of transitions can be supported by increasing the expressiveness of the Gemini grammar. 
Recommendations could also be made finer-grained; for example, data changes can be divided into more specific units (e.g., aggregation, binning, filtering) by examining the data transformations of the start and end visualization states. Gemini could then apply more detailed guidelines for evaluation.
Moreover, knowledge-based recommendation methods~\cite{draco, dziban} are more scalable in terms of handling evaluation factors (or changes) than our current heuristic methods. Open challenges include how to further formalize animation guidelines~\cite{anim_transition, tversky} and our complexity measure within constraint programming systems.

Finally, to obtain more elaborate staged animations, it seems promising to recommend intermediate visualization states (``keyframes'') and then cascade Gemini animations. For example, when points move from a scale domain of $[0,1]$ to $[11,12]$, it might be preferable to designate an intermediate state using a domain of $[0,12]$ before moving the points, letting viewers see the 10-plus-length movements of the points. One way to recommend intermediate states is to leverage GraphScape~\cite{graphscape} to explore and rank transition paths. After selecting a transition path with proper intermediate visualization states, Gemini could generate animated transitions for each hop (transition) on the path, then combine them into a final animation. For instance, the transition mentioned earlier ($S \xrightarrow{} E$) can be divided into three Gemini animated transitions with a path including two intermediate states: ($S \xrightarrow{} S_{[0,12]}$), ($S_{[0,12]} \xrightarrow{} E_{[0,12]}$), ($E_{[0,12]} \xrightarrow{} E$), where $S_{[0,12]}$ and $E_{[0,12]}$ are the start and end states on a $[0,12]$ domain. With such extensions, Gemini could help facilitate a mixed-initiative keyframe authoring paradigm~\cite{anim_authoring_env}.


As noted, the expressiveness of Gemini is limited to transitions among single view charts. To express transitions for more varied charts (e.g., multi-view, pie, trail, and geographic charts), the grammar needs to be extended to support new components such as headers and mark groups, reference geographic projections, and support shape interpolation (e.g., pie to bar). In addition, the Gemini grammar cannot assign separate timings across changes within a step (as we failed to replicate for P4-3). 
One way to support such nuanced timing is to allow the grammar to assign separate timings for low-level component properties (e.g., individual timings for each scale for a mark).

Regarding the Gemini implementation, remaining challenges involve the computational complexity stemming from data and step count sizes. The compiler and recommender system join data and enumeration sequences that combinatorially increase with the number of stages. 
For example, on the first author's personal laptop, compiling a Gemini spec for random movements of 5,000 points took 300-500 ms (versus 10-60 ms in D3), and the compiled animation played 300-600 ms (0-60 ms in D3) longer than the specified duration (5000 ms).
In addition, the recommendation feature takes more than one minute to produce recommendations with $N=4$ stages. Future compiler optimizations\,---\,such as indexing for data joins and dynamic programming to enumerate and evaluate the timelines\,---\,can improve the performance.



\vspace{-0.1cm}
\acknowledgments{
We thank Arvind Satyanarayan, Dominik Moritz, \& Kanit Wongsuphasawat for their feedback. NSF award IIS-1907399 supported this work.}

\bibliographystyle{abbrv-doi}

\bibliography{reference}
\end{document}